\documentstyle[12pt]{article}
\title{Super--energy tensor for space--times with
vanishing scalar curvature}

\author{Miguel \'A.G. Bonilla\footnotemark[3] \footnotemark[1]
\, and Carlos F. Sopuerta\footnotemark[4] \thanks{Also 
at Laboratori de F\'{\i}sica Matem\`atica,
Societat Catalana de F\'{\i}sica, IEC, Barcelona.  
E--mail: mangel@ffn.ub.es, cfs@tpi.uni--jena.de}  \\
\ddag\ Departament de F\'{\i}sica Fonamental,
Universitat de Barcelona,\\
Avda. Diagonal 647, E--08028 Barcelona, Spain \\
$\S$\ Institut for Theoretical Physics, FSU Jena, \\
Max--Wien--Platz 1, D--07743 Jena, Germany}

\begin{document}

\renewcommand{\thesection}{\Roman{section}}

\maketitle

\begin{abstract}
A four--index tensor is constructed with terms
both quadratic in the Riemann tensor and
linear in its second derivatives,
which has zero divergence for space--times
with vanishing scalar curvature.
This tensor reduces in vacuum to the Bel--Robinson
tensor. Furthermore, the completely timelike component
referred to any observer is positive, and zero if and
only if the space--time is flat (excluding some
unphysical space--times). We also show that this
tensor is the unique that can be constructed
with these properties. Such a tensor does
not exist for general gravitational fields.
Finally, we study this tensor in several examples:
the Friedmann--Lema\^{\i}tre--Robertson--Walker space--times
filled with radiation, the plane--fronted gravitational 
waves, and the Vaidya radiating metric.
\end{abstract}


\section{Introduction}
The investigation of conservation laws in general
relativity has a long history. From its very
beginning much of this research was based on
pseudo--tensors instead of fully covariant
methods. The aim of many of these works was
to find differential laws which, once transformed
into integral ones, were interpreted as
energy balances in such a way that expressions
for the energy and momentum densities of the
gravitational field could be identified.

The covariant approach to this problem is mainly based
on the analogy of the Bel--Robinson tensor \cite{BEL1} with
the energy--momentum tensor of the electromagnetic
field (there are other approaches based on this analogy, see
for instance~\cite{MMCQ}). 
The Bel--Robinson tensor is conserved
{\em in vacuum}, completely symmetric and traceless.
Moreover, the completely timelike component referred
to any observer (described by a timelike unit
vector field) is non--negative, and its vanishing
implies that the space--time is conformally flat (flat
in vacuum). This is a desirable positivity property
for any candidate to gravitational energy density.
In spite of these good properties, the Bel--Robinson
tensor has dimensions of energy density square
and this fact makes its interpretation somewhat unclear. 
Nevertheless, it has revealed as a very useful
tool in many kinds of studies, which has led to
some efforts in finding extensions of the Bel--Robinson
tensor for more general cases than vacuum.
Therefore, the question arises whether
generalizations of the Bel--Robinson tensor exist
for space--times not necessarily empty. 

The Bel tensor \cite{BEL2} was the first attempt on this
problem. It is a tensor whose completely timelike component
is positive and zero only when the space--time is
Minkowski. In vacuum, it reduces to the Bel--Robinson
tensor, but in the general case it is no longer conserved.

For general space--times, Sachs \cite{SACH}
found a divergence--free tensor that coincides with the
Bel--Robinson tensor in vacuum. Unfortunately, this tensor
does not satisfy any positivity property
and it is neither completely symmetric nor traceless.

The systematic treatment of this problem was made
by Collinson \cite{COLL}, who found all four--index
divergence--free tensors with terms either
quadratic in the Riemann tensor or linear in its
second derivatives. The result found out is that
any such tensor can be derived from only {\em one}
tensor, namely $T_{10}^{\alpha\beta\lambda\mu}$ (see
(\ref{colli}) in the appendix A), whose divergence with
respect to the first index vanishes.
However, there is not enough freedom to construct
a tensor with its time component positive.

In this paper we show that, unlike the general case,
for space--times with zero scalar curvature ($R=0$)
it is possible to construct a unique generalization
of the Bel--Robinson tensor.
We begin in Sect. 2 by proving that, when $R=0$,
there exists another conserved tensor which cannot be
derived from the Collinson one. In Sect. 3 we show that,
demanding symmetry in the three free indices,
there is not any other tensor independent
from these two. This new tensor allows us to construct
(Sect. 4) a divergence--free tensor which has the
completely timelike component non--negative and
zero only when the space--time is flat (excluding
some cases that, via Einstein's field equations, have
an unphysical matter content). This tensor is completely
symmetric in its three last indices, but it is impossible
to get a similar tensor symmetric in all their indices.
We remark that it is not possible to construct any
other tensor with such characteristics.  

In order to illustrate this development, we study in 
Sect.~5 some examples in which we can define this tensor,
namely: the Friedmann--Lema\^{\i}tre--Robertson--Walker
(FLRW) models with a energy--momentum content of (incoherent)
radiation ($p=\varrho/3$), the plane--fronted gravitational 
waves with parallel rays ({\em pp waves}), and the Vaidya 
radiating space--time. 

Finally, we recall that for purely 
electromagnetic space--times, and supposing that the 
Einstein field equations hold, Penrose and Rindler 
\cite{PERI} also gave a generalization of the Bel--Robinson 
tensor by using spinor methods. This tensor is conserved,
completely symmetric and traceless in its three last
indices. In the appendix B we find its tensorial
expression and a new (up to our knowledge) positivity
property.

\section{Deduction of the new conserved tensor for $R=0$}
Otherwise said, throughout this paper we will consider
the metric tensor $g_{\alpha\beta}$ to have signature
$(-,+,+,+)$. The convention for indices on the Riemann
tensor that will be used is defined through the Ricci identities:
\begin{eqnarray}
\left(\nabla_{\alpha}\nabla_{\beta}-
\nabla_{\beta}\nabla_{\alpha}\right)v_{\lambda}=
-R^{\sigma}_{\hspace{1mm}\lambda\alpha\beta}v_{\sigma} \, ,
\label{ricci}
\end{eqnarray}
where $v_{\alpha}$ is an arbitrary 1--form. The Ricci tensor
and the scalar curvature are defined as usual:
$R_{\alpha\beta}\equiv
R^{\sigma}_{\hspace{1mm}\alpha\sigma\beta}$ and
$R\equiv R^{\sigma}_{\sigma}$. We also recall
the Riemann symmetries and the first and second
Bianchi identities:
\begin{eqnarray}
& R_{\alpha\beta\lambda\mu}=R_{[\alpha\beta][\lambda\mu]}
=R_{\lambda\mu\alpha\beta}\, ,
\nonumber \\
& \hspace{1cm} R_{[\alpha\beta\lambda]\mu}=0 \, ,  
\nonumber\\
& \nabla_{[\nu}R_{\alpha\beta]\lambda\mu}=0 \, .
\label{bianchi}
\end{eqnarray}

The procedure we are going to use to find the conserved tensor
starts from the expression for the divergence of
the Bel tensor \cite{BEL1}:
\begin{eqnarray}
\nabla_{\alpha}T^{\alpha\beta\lambda\mu}=
R^{\beta\hspace{1.5mm}\lambda}_{\hspace{1mm}\rho
\hspace{2mm}\sigma}J^{\mu\sigma\rho}+
R^{\beta\hspace{1.5mm}\mu}_{\hspace{1mm}\rho
\hspace{2mm}\sigma}J^{\lambda\sigma\rho}
-\frac{1}{2}g^{\lambda\mu}R^{\beta}_{\hspace{1mm}
\rho\sigma\gamma}J^{\sigma\gamma\rho} \, ,
\label{divbel}
\end{eqnarray}
where the Bel tensor $T^{\alpha\beta\lambda\mu}$
and $J^{\alpha\beta\lambda}$ are defined as follows:
\begin{eqnarray}
T^{\alpha\beta\lambda\mu}&\equiv&
\frac{1}{2}\left(R^{\alpha\rho\lambda\sigma}  
R^{\beta\hspace{1.5mm}\mu}_{\hspace{1mm}
\rho\hspace{2mm}\sigma}+   
{*R*}^{\alpha\rho\lambda\sigma}
{*R*}^{\beta\hspace{1.5mm}\mu}_{\hspace{1mm}
\rho\hspace{2mm}\sigma}+   
{*R}^{\alpha\rho\lambda\sigma}
{*R}^{\beta\hspace{1.5mm}\mu}_{\hspace{1mm}
\rho\hspace{2mm}\sigma}+
{R*}^{\alpha\rho\lambda\sigma}
{R*}^{\beta\hspace{1.5mm}\mu}_{\hspace{1mm}
\rho\hspace{2mm}\sigma}\right) \, ,
\nonumber\\
J^{\lambda\mu\beta}&\equiv&
\nabla^{\lambda}R^{\mu\beta}-
\nabla^{\mu}R^{\lambda\beta}=
\nabla_{\sigma}R^{\mu\lambda\beta\sigma} \, ,
\label{J}
\end{eqnarray}
being ``$\ast$'' the usual dual operator acting over any
pair of antisymmetric indices:
\begin{eqnarray*}
& {*R}_{\alpha\beta\lambda\mu}\equiv
\frac{1}{2} \eta_{\alpha\beta\sigma\rho}
R^{\sigma\rho}_{\hspace{3mm}\lambda\mu} \, , \;\;
{R*}_{\alpha\beta\lambda\mu}\equiv
\frac{1}{2} \eta_{\lambda\mu\sigma\rho}
R_{\alpha\beta}^{\hspace{3mm}\sigma\rho} \, , \\
& {*R*}_{\alpha\beta\lambda\mu}\equiv
\frac{1}{4} \eta_{\alpha\beta\gamma\delta}
\eta_{\lambda\mu\sigma\rho}
R^{\gamma\delta\sigma\rho} \, ,
\end{eqnarray*}
and $\eta_{\alpha\beta\lambda\mu}$ is the canonical volume
4--form. Our purpose now is to work out the right hand side (r.h.s.)
of equation (\ref{divbel}) in order to convert it into a global
divergence. To that end, we will repeatedly integrate by parts
and make use of equations (\ref{ricci},\ref{bianchi}).

To begin with, the last term in equation (\ref{divbel}) can be
easily transformed into a divergence, by means of the Ricci
and Bianchi identities (\ref{ricci},\ref{bianchi}):
\begin{eqnarray}
-\frac{1}{2}g^{\lambda\mu}R^{\beta}_{\hspace{1mm}
\rho\sigma\gamma}J^{\sigma\gamma\rho}=
-g^{\lambda\mu}\nabla_{\alpha}\nabla_{\rho}
J^{\alpha\rho\beta} \, .
\label{1term}
\end{eqnarray}
Next we expand the leading terms of the r.h.s. of
(\ref{divbel}) by using the definition of
$J^{\alpha\beta\lambda}$ (\ref{J}) and integrating by parts:
\begin{eqnarray}
R^{\beta\hspace{1.5mm}\lambda}_{\hspace{1mm}\rho
\hspace{2mm}\sigma}\nabla^{\mu}R^{\sigma\rho}-
\nabla_{\alpha}\left(R^{\mu\rho}
R^{\beta\hspace{2mm}\lambda\alpha}_{
\hspace{1mm}\rho}\right)+
R^{\mu\rho}\nabla_{\alpha}
R^{\beta\hspace{2mm}\lambda\alpha}_{
\hspace{1mm}\rho}+
\left[\lambda \longleftrightarrow \mu\right] \, .
\label{2term}
\end{eqnarray}
The first term in the previous expression can be
rewritten by means of the Ricci identities (\ref{ricci})
as follows:
\begin{eqnarray}
R^{\beta\hspace{1.5mm}\lambda}_{\hspace{1mm}\rho
\hspace{2mm}\sigma}\nabla^{\mu}R^{\sigma\rho}=
\left(\nabla^{\lambda}\nabla_{\sigma}-
\nabla_{\sigma}\nabla^{\lambda}\right)
\nabla^{\mu}R^{\sigma\beta}+ 
\nonumber\\
+\left[-\nabla^{\alpha}\left(
R^{\mu\hspace{1.5mm}\lambda}_{\hspace{1mm}\alpha
\hspace{2mm}\sigma}R^{\sigma\beta}\right)+
R^{\sigma\beta}\nabla^{\alpha}
R^{\lambda\hspace{1.5mm}\mu}_{\hspace{1mm}\sigma
\hspace{2mm}\alpha}\right]
+R_{\sigma}^{\lambda}\nabla^{\mu}R^{\sigma\beta} \, .
\label{3term}
\end{eqnarray}
Thus, we have converted the r.h.s. of equation (\ref{divbel})
into a divergence plus the following terms:
\begin{eqnarray}
R^{\sigma\beta}\nabla^{\alpha}
R^{\lambda\hspace{1.5mm}\mu}_{\hspace{1mm}\sigma
\hspace{2mm}\alpha}+
R^{\sigma\lambda}\nabla^{\mu}R_{\sigma}^{\beta}+
R^{\mu\rho}\nabla_{\alpha}
R^{\beta\hspace{2mm}\lambda\alpha}_{
\hspace{1mm}\rho}+
\left[\lambda \longleftrightarrow \mu\right]=
\nonumber \\
=R^{\sigma\beta}\left(\nabla_{\sigma}R^{\lambda\mu}
-\nabla^{\lambda}R_{\sigma}^{\mu}\right)
+R^{\sigma\lambda}\nabla^{\mu}R_{\sigma}^{\beta}+
R^{\mu\rho}\left(\nabla_{\rho}R^{\beta\lambda}
-\nabla^{\beta}R_{\rho}^{\lambda}\right)
+\left[\lambda \longleftrightarrow \mu\right] \, .
\label{4term}
\end{eqnarray}
Now, taking into account the contracted
Bianchi identities ($\nabla_{\mu}R^{\mu\nu}=
\frac{1}{2}\nabla^{\nu}R$), these terms can be transformed
into the following expression:
\begin{eqnarray}
-\nabla^{\beta}\left(R_{\sigma}^{\lambda}
R^{\sigma\mu}\right)+
\nabla^{\mu}\left(R_{\sigma}^{\lambda}
R^{\sigma\beta}\right)+
\nabla^{\lambda}\left(R_{\sigma}^{\mu}
R^{\sigma\beta}\right)
-2 R^{\beta\sigma}\left(\nabla^{\mu}R_{\sigma}^{\lambda}+
\nabla^{\lambda}R_{\sigma}^{\mu}\right)+
\nonumber\\
+\nabla_{\sigma}\left(R^{\beta\lambda}R^{\mu\sigma}+
R^{\beta\mu}R^{\lambda\rho}+
2 R^{\beta\sigma}R^{\lambda\mu}\right)-
\frac{1}{2}\left(R^{\beta\lambda}\nabla^{\mu}R+
R^{\beta\mu}\nabla^{\lambda}R+
2 R^{\lambda\mu}\nabla^{\beta}R\right) \, .
\label{5term}
\end{eqnarray}
The last three terms of this expression vanish when
$R$ is constant, so we are finally left with $-2R^{\beta\sigma}
\left(\nabla^{\mu}R_{\sigma}^{\lambda}+
\nabla^{\lambda}R_{\sigma}^{\mu}\right)$. Nevertheless,
notice that our final purpose is to find
a conserved tensor ${T''}^{\alpha\beta\lambda\mu}$
whose completely timelike component referred to an
observer $\vec{u}$, ${T''}^{\alpha\beta\lambda\mu} 
u_{\alpha}u_{\beta}u_{\lambda}u_{\mu}$, is positive,
which means that we are only interested in the symmetric
part. Therefore, without lost of generality, we can
symmetrize the whole expression and, as a consequence,
the remaining terms transform themselves into a divergence: 
\begin{eqnarray}
-4R^{\sigma(\beta}
\nabla^{\lambda}R_{\sigma}^{\mu)}=
-2\nabla^{(\beta}\left(R_{\sigma}^{\lambda}
R^{\mu)\sigma}\right) \, .
\label{6term}
\end{eqnarray}

So, we have finally achieved a conserved tensor
if the scalar curvature vanishes (in fact, if
it is constant). Collecting all the
previous terms (\ref{1term}-\ref{6term}) we get the
final result:
\begin{eqnarray*}
\nabla_{\alpha}{T''}^{\alpha\beta\lambda\mu}=
-2 R^{(\beta\lambda}\nabla^{\mu)}R \, ,
\end{eqnarray*}
where we have defined
\begin{eqnarray}
{T''}^{\alpha\beta\lambda\mu}&\equiv&
{T}^{\alpha(\beta\lambda\mu)}-
4 R^{\alpha(\beta}R^{\lambda\mu)}+
g^{\alpha(\beta}R^{\lambda}_{\sigma}R^{\mu)\sigma}-
\nonumber\\
&-& 2\nabla^{(\beta}\nabla^{\lambda}R^{\mu)\alpha}+
2\nabla^{(\beta}\nabla^{|\alpha|}R^{\lambda\mu)}-
2\nabla^{\alpha}\nabla^{(\beta}R^{\lambda\mu)}-
\nonumber\\
&-& 2 g^{\alpha(\beta}\nabla_{\sigma}\nabla^{\lambda}
R^{\mu)\sigma}-
\nabla_{\sigma}\nabla^{\sigma}R^{\alpha(\beta}
g^{\lambda\mu)}+
\nabla_{\sigma}\nabla^{\alpha}R^{\sigma(\beta}
g^{\lambda\mu)} \, .
\label{T''}
\end{eqnarray}
It is a matter of checking that this tensor cannot
be obtained from Collinson tensor
$T_{10}^{\alpha\beta\lambda\mu}$
when it is restricted to the case $R=0$.
Now, we are left with the question of its uniqueness.

\section{Uniqueness}
In this section we will prove that, in the case
we are concerned with ($R=0$), it does not exist
any other conserved tensor, symmetric in its
three last indices, independent from
${T''}^{\alpha\beta\lambda\mu}$ and the Collinson
tensor, ${T}_{10}^{\alpha\beta\lambda\mu}$.
The reasoning is the following. Suppose you are given
a tensor ${T'''}^{\alpha\beta\lambda\mu}$ which is conserved
when $R=0$. The divergence computed in the general case
will be a combination of the following type (taking into
account symmetries and unit dimensions):
\begin{eqnarray*}
\nabla_{\alpha}{T'''}^{\alpha\beta\lambda\mu}=
a R^{(\beta\lambda}\nabla^{\mu)}R+
b R\nabla^{(\beta}R^{\lambda\mu)}+
c R^{\sigma(\beta}\nabla_{\sigma}R g^{\lambda\mu)}+
d R g^{(\beta\lambda}\nabla^{\mu)}R+
\\
+ e \nabla^{(\beta}\nabla^{\lambda}\nabla^{\mu)}R+
f \nabla^{\sigma}\nabla_{\sigma}\nabla^{(\beta}R
g^{\lambda\mu)}+
h \nabla^{\sigma}\nabla^{(\beta}\nabla_{\sigma}R
g^{\lambda\mu)}+
i \nabla^{(\beta}\nabla^{|\sigma|}\nabla_{\sigma}R
g^{\lambda\mu)} \, ,
\end{eqnarray*}
$a$, $b$, $c$, $d$, $e$, $f$, $h$ and $i$ being constants.
This can be immediately cast in the following form:
\begin{eqnarray*}
\nabla_{\alpha}{T'''}^{\alpha\beta\lambda\mu}=
(a-b)R^{(\beta\lambda}\nabla^{\mu)}R+
\nabla_{\alpha}\tau^{\alpha\beta\lambda\mu} \, ,
\end{eqnarray*}
where $\tau^{\alpha\beta\lambda\mu}$ stands for:
\begin{eqnarray*}
\tau^{\alpha\beta\lambda\mu}\equiv
b R g^{\alpha(\beta}R^{\lambda\mu)}+
c \left(R R^{\alpha(\beta}g^{\lambda\mu)}-
\frac{1}{4} g^{\alpha(\beta}g^{\lambda\mu)}R^{2}\right)+
d \frac{1}{2} g^{(\beta\lambda}g^{\mu)\alpha}R^{2}+ \\
+ e g^{\alpha(\beta}\nabla^{\lambda}\nabla^{\mu)}R+
f g^{\alpha\sigma}\nabla_{\sigma}\nabla^{(\beta}R
g^{\lambda\mu)}+
h g^{\alpha\sigma}\nabla^{(\beta}\nabla_{\sigma}R
g^{\lambda\mu)}+
i g^{\alpha(\beta}\nabla^{|\sigma|}\nabla_{\sigma}R
g^{\lambda\mu)} \, ,
\end{eqnarray*}
so clearly it is a tensor that vanishes when $R$ does.
That is, if a tensor of the kind we are considering
is divergence--free when $R=0$, in the 
general case its divergence should be a multiple of
$R^{(\beta\lambda}\nabla^{\mu)}R$ plus the divergence
of a tensor of the type $\tau^{\alpha\beta\lambda\mu}$.
If we had two such tensors, a suitable combination
of them removing the term $R^{(\beta\lambda}\nabla^{\mu)}R$
would give a conserved tensor for the general
case and, therefore, due to the Collinson result \cite{COLL},
it could be constructed from
$T_{10}^{\alpha\beta\lambda\mu}$. Given that
the tensor $\tau^{\alpha\beta\lambda\mu}$
vanishes when $R=0$, the three tensors would not
be independent in that case.

On the other hand, this reasoning shows that, from the
very beginning, we were able to know that in the $R=0$
case at most one more conserved tensor could exist apart
from Collinson's one, as finally has been the case.

\section{Positivity}
As it has been pointed out above, in the general case
all the conserved tensors can be constructed from
$T_{10}^{\alpha\beta\lambda\mu}$. This construction is based
in two procedures. First, it is clear that if we perform any
permutation on the three last indices we will still have a
conserved tensor. Second, by taking the two traces 
$T_{10\hspace{2.5mm}\rho}^{\alpha\beta\rho}$,
$T_{10\hspace{2.5mm}\rho}^{\alpha\rho\beta}$
and multiplying them by $g^{\lambda\mu}$ we obtain new
conserved tensors. Actually, there is not any other
two--index divergence--free tensor independent from
them. These two tensors can be taken to be (as usually
obtained by hamiltonian differentiation):
\begin{eqnarray*}
t_{1}^{\alpha\beta}&=&2\nabla^{\alpha}\nabla^{\beta}R-
2g^{\alpha\beta}\nabla_{\mu}\nabla^{\mu}R+
\frac{1}{2}g^{\alpha\beta}R^{2}-
2 R R^{\alpha\beta} \, , \\
t_{2}^{\alpha\beta}&=&
2\nabla_{\sigma}\nabla^{\alpha}R^{\sigma\beta}-
\nabla_{\sigma}\nabla^{\sigma}R^{\alpha\beta}-
2R^{\alpha\sigma}R^{\beta}_{\sigma}+
\frac{1}{2}R_{\sigma\rho}R^{\sigma\rho}g^{\alpha\beta}-
\frac{1}{2}g^{\alpha\beta}\nabla_{\mu}\nabla^{\mu}R \, . 
\end{eqnarray*}
This is all the freedom we have in the general case.

In the $R=0$ case, there exists only one two--index
conserved tensor, which is that obtained from
$t_{2}^{\alpha\beta}$. On the other hand, as we consider
tensors which are symmetric in the three last indices,
from the Collinson tensor we will only have one
independent tensor (apart from the two--index tensor),
namely $T_{10}^{\alpha(\beta\lambda\mu)}$ or, equivalently,
the Sachs tensor $T'^{\alpha\beta\lambda\mu}$ \cite{SACH}
which is symmetric in its three last indices.

Therefore, to construct the super--energy tensor
in the $R=0$ case, we are led with three tensors:
{\em i)} the Sachs tensor $T'^{\alpha\beta\lambda\mu}$
(restricted to the $R=0$ case):
\begin{eqnarray*}
{T'}^{\alpha\beta\lambda\mu}&\equiv&
{T}^{\alpha(\beta\lambda\mu)}+
2 R^{\alpha\sigma}R^{(\beta}_{\sigma}g^{\lambda\mu)}
-\frac{1}{3}
g^{\alpha(\beta}R^{\lambda}_{\sigma}R^{\mu)\sigma}
-\frac{1}{2}
R_{\sigma\rho}R^{\sigma\rho}
g^{\alpha(\beta}g^{\lambda\mu)}+\\
&+&2\nabla^{(\beta}\nabla^{\lambda}R^{\mu)\alpha}-
\frac{2}{3}\nabla^{(\beta}\nabla^{|\alpha|}R^{\lambda\mu)}-
\frac{2}{3}\nabla^{\alpha}\nabla^{(\beta}R^{\lambda\mu)}+\\
&+&\frac{4}{3}\nabla_{\sigma}\nabla^{\sigma}R^{(\beta\lambda}
g^{\mu)\alpha}
-2 g^{\alpha(\beta}
\nabla_{\sigma}\nabla^{\lambda}R^{\mu)\sigma}-
\nabla_{\sigma}\nabla^{\alpha}R^{\sigma(\beta}
g^{\lambda\mu)} \, ,
\end{eqnarray*}
{\em ii)} the two--index tensor $t^{\alpha\beta}$:
\begin{eqnarray*}
t^{\alpha\beta}=2\nabla_{\sigma}\nabla^{\alpha}R^{\sigma\beta}-
\nabla_{\sigma}\nabla^{\sigma}R^{\alpha\beta}-
2R^{\alpha\sigma}R^{\beta}_{\sigma}+
\frac{1}{2}R_{\sigma\rho}R^{\sigma\rho}g^{\alpha\beta} \, ,
\end{eqnarray*}
and {\em iii)} the tensor $T''^{\alpha\beta\lambda\mu}$
previously found in (\ref{T''}).

Now, we have to combine these three tensors in such a way
that any observer measures a positive quantity. Moreover,
we would like that this completely timelike component
vanishes if and only if the space--time is flat.

First of all, we have to take into account that terms
made of derivatives of the Ricci tensor do not have a definite
sign, so it would be necessary to eliminate their
contributions. This aim can only be achieved by means of the
following combination:
\begin{eqnarray*}
A^{\alpha\beta\lambda\mu}\equiv \frac{1}{2}\left(
3{T'}^{\alpha\beta\lambda\mu}-
{T''}^{\alpha\beta\lambda\mu}\right)+
\frac{5}{2}t^{\alpha(\beta} g^{\lambda\mu)}\, , 
\end{eqnarray*}
which more explicitly reads:
\begin{eqnarray}
A^{\alpha\beta\lambda\mu}&=&T^{\alpha(\beta\lambda\mu)}+
\nonumber \\
&+&2R^{\alpha(\beta}R^{\lambda\mu)}-
2R^{\alpha\sigma}R^{(\beta}_{\sigma}g^{\lambda\mu)}-
g^{\alpha(\beta}R^{\lambda}_{\sigma}R^{\mu)\sigma}+
\frac{1}{2}R_{\sigma\rho}R^{\sigma\rho}
g^{\alpha(\beta}g^{\lambda\mu)}+ \nonumber \\
&+&2\nabla^{(\beta}\nabla^{\lambda}R^{\mu)\alpha}-
2\nabla^{(\beta}\nabla^{|\alpha|}R^{\lambda\mu)}
+3\nabla_{\sigma}\nabla^{\alpha}R^{\sigma(\beta}g^{\lambda\mu)}-
\nonumber \\
&-&2g^{\alpha(\beta}\nabla_{\sigma}\nabla^{\lambda}R^{\mu)\sigma}+
2\nabla_{\sigma}\nabla^{\sigma}R^{(\beta\lambda}g^{\mu)\alpha}-
2\nabla_{\sigma}\nabla^{\sigma}R^{\alpha(\beta}g^{\lambda\mu)} 
\,. \label{tena}
\end{eqnarray}
Since we have already exhausted all the freedom, we
finally examine the completely timelike component
referred to any timelike unit vector $\vec{u}$:
\begin{eqnarray}
A\left(\vec{u}\right)&\equiv&
A_{\alpha\beta\lambda\mu}u^{\alpha}u^{\beta}u^{\lambda}u^{\mu}=
T_{\alpha\beta\lambda\mu}u^{\alpha}u^{\beta}u^{\lambda}u^{\mu}+
\frac{1}{2}R_{\sigma\rho}R^{\sigma\rho}+ 
\nonumber \\
&+&2\left(R_{\alpha\beta}u^{\alpha}u^{\beta}\right)^{2}+
3\left(R^{\alpha\sigma}R^{\beta}_{\sigma}\right)
u_{\alpha}u_{\beta}-
\left(\nabla_{\sigma}\nabla^{\alpha}R^{\sigma\beta}\right)
u_{\alpha}u_{\beta} \, .
\label{a}
\end{eqnarray}
To check the positivity of $A\left(\vec{u}\right)$ it is
convenient to write out the last term of the previous
expression in the following form:
\begin{eqnarray*}
\nabla_{\sigma}\nabla^{\alpha}R^{\sigma\beta}=
C^{\alpha\hspace{3mm}\beta}_{\hspace{2mm}\sigma\rho}
R^{\sigma\rho}+
2R^{\alpha\sigma}R^{\beta}_{\sigma}-
\frac{1}{2}g^{\alpha\beta}R_{\sigma\rho}R^{\sigma\rho} \, .
\end{eqnarray*} 
At this point, we introduce four spatial tensors, namely
$E_{\alpha\beta}(\vec{u})$, $H_{\alpha\beta}(\vec{u})$,
$M_{\alpha\beta}(\vec{u})$ and $N_{\alpha\beta}(\vec{u})$,
that (together with $R$) wholly characterize the Riemann
tensor \cite{BOSE}. Their definitions, properties and some
useful formulae are given in the appendix A.

Introducing the previous definitions in (\ref{a})
we obtain, after some calculations:
\begin{eqnarray}
A\left(\vec{u}\right)=
\left(E_{\sigma\rho}-M_{\sigma\rho}\right)
\left(E^{\sigma\rho}-M^{\sigma\rho}\right)+
H_{\sigma\rho}H^{\sigma\rho}+
3N_{\sigma\rho}N^{\sigma\rho}+
\left(R_{\alpha\beta}u^{\alpha}u^{\beta}\right)^{2} \,,
\label{comt}
\end{eqnarray}
which is a sum of square terms (all the tensors appearing
here are spatial). Therefore it is manifestly positive and its
vanishing implies:
\begin{eqnarray*}
H_{\alpha\beta}=0\, , \; \; N_{\alpha\beta}=0 \, ,\\
R_{\alpha\beta}u^{\alpha}u^{\beta}=0 \, , \; \;
E_{\alpha\beta}=M_{\alpha\beta} \, .
\end{eqnarray*}
The previous expressions lead to
$R_{\alpha\beta}u^{\alpha}=0$ (see appendix A).
This condition, when considering the Einstein field
equations, immediately drives to an unphysical
energy--momentum tensor. Hence, if we eliminate these
unphysical space--times (for instance, adding any energy
condition), the vanishing of $A\left(\vec{u}\right)$
finally implies the Minkowski space--time.

\section{Some examples}

In this section we are going to study the tensor 
$A^{\alpha\beta\lambda\mu}$ (\ref{tena}) in some space--times 
with vanishing scalar curvature.  In particular, we are going 
to consider the following examples: 
(i) the radiation FLRW cosmological models, (ii) the 
{\em pp waves}, and (iii) the Vaidya radiating metric.
In examples (i) and (ii) we will give the expression for 
$A^{\alpha\beta\lambda\mu}$ and its completely timelike
component $A(\vec{u})$ (\ref{comt}) for an arbitrary observer 
$\vec{u}$. In the example (iii), for the sake of brevity we 
will give only the expression of $A(\vec{u})$, also for an 
arbitrary observer.

\vspace{4mm}

In the first example we study the case of the FLRW models
(see for instance \cite{KSHM}) with vanishing scalar curvature, 
the radiation models, whose
energy--momentum tensor (of perfect--fluid type) is given by the 
following expression (throughout this section we will use units 
in which $8\pi G=c=1$)
\[ T_{\alpha\beta} = \varrho \,U_\alpha U_\beta + p\,
h_{\alpha\beta}\,, \hspace{14mm} p=\frac{1}{3}\varrho \,,\]
where $\vec{U}$ is the fluid velocity ($U^\alpha U_\alpha=-1$),
$\varrho$ the energy density, $p$ the pressure, and 
$h_{\alpha\beta}=g_{\alpha\beta}+U_\alpha U_\beta$ the 
orthogonal projector to the fluid velocity.  The line element of 
these conformally--flat models can be written as 
\[ ds^2=-dt^2+a^2(t)\left\{d\chi^2+\Sigma^2(\epsilon,\chi)
\,(d\theta^2+\sin^2\theta d\varphi^2)\right\}\,, \label{lerw}\] 
where $\Sigma(\epsilon,\chi)$  is given by
\[ \Sigma(\epsilon,\chi) = \left\{ \begin{array}{ll} 
\sin\chi  & \mbox{if $\epsilon = 1$\, ,}  \\ 
\chi   & \mbox{if $\epsilon = 0$\, ,}  \\ 
\sinh\chi & \mbox{if $\epsilon = -1$\,.}\end{array}\right.\] 
The fluid velocity $\vec{U}$, the scale factor $a(t)$ and the energy 
density $\varrho(t)$ are
\[\vec{U}=\frac{\partial}{\partial t}\,, \hspace{8mm} 
a^2(t)= (t-t_o)\left[2A-\epsilon(t-t_o)\right] \,, 
\hspace{8mm} \varrho(t) = \frac{3A^2}{a^4(t)} \,,\]
respectively, and $A$ and $t_o$ are arbitrary constants.

After some straightforward calculations, and using the 
special properties of these space--times, we arrive at the 
following expression for $A^{\alpha\beta\lambda\mu}$
\begin{equation}
A^{\alpha\beta\lambda\mu} = 
\frac{4}{3}\varrho^2\left\{ U^\alpha U^\beta U^\lambda
U^\mu + U^\alpha U^{(\beta} h^{\lambda\mu)} +
\frac{5}{3}h^{\alpha(\beta}U^\lambda U^{\mu)}+
\frac{1}{3}h^{\alpha(\beta}h^{\lambda\mu)}\right\}\,. 
\label{afrw}
\end{equation}
As we can see, it is proportional to the energy density squared.
We can also check that it is indeed divergence--free.  Now, let 
us compute the completely timelike component (\ref{comt}) of this 
tensor with respect to an arbitrary observer $\vec{u}$.  To 
that end, we decompose $\vec{u}$ in the next way
\[ \vec{u} = \gamma(\vec{U}+\vec{v})\,, \hspace{6mm}
v^\alpha U_\alpha =0\,, \hspace{4mm} 
v^\alpha v_\alpha = v^2 \geq 0\,, \hspace{4mm}
\gamma\equiv (1-v^2)^{-1/2} \,, \]
where the case $\vec{v}=0$ corresponds to an observer comoving 
with the fluid ($\vec{u}=\vec{U}$). Then, from 
(\ref{comt},\ref{afrw}) 
we find that $A(\vec{u})$ is given by
\[ A(\vec{u}) = \frac{4}{3}\varrho^2\gamma^4\left\{
1+\frac{8}{3}v^2+\frac{1}{3}v^4\right\} \,. \]
That is, it is function of $\varrho$ and $v$ only.  Moreover,
it increases monotonically as $v$ increases and its minimum
corresponds to the case $v=0$, in which the observer is 
comoving with the fluid.

\vspace{4mm}

Now, we are going to study the tensor 
$A^{\alpha\beta\lambda\mu}$ in the case of
the {\em pp waves} space--times.  The corresponding
line element can be written in  null coordinates 
$\{u,v,\zeta,\bar{\zeta}\}$ as follows (see \cite{KSHM}
for details)
\[ ds^2 = -2 du dv + 2 d\zeta d\bar{\zeta} -2 H du^2\,,\]
where $H$ is an arbitrary function which does not depend on $v$ 
[$H=H(u,\zeta,\bar{\zeta})$]. Using the following Newman--Penrose
basis $\{\ell,k,m,\bar{m}\}$
\[ \vec{\ell}=\frac{\partial}{\partial v}\,,\hspace{5mm}
\vec{k} = \frac{\partial}{\partial u}-H
\frac{\partial}{\partial v}\,,\hspace{5mm}
\vec{m}= \frac{\partial}{\partial \zeta}\,, \]
the Ricci and self--dual Weyl tensors are 
\begin{equation}
R_{\alpha\beta}=2\Phi\ell_\alpha\ell_\beta\,,\hspace{1cm}
\hat{C}_{\alpha\beta\lambda\mu}\equiv 
C_{\alpha\beta\lambda\mu}+i\stackrel{*}{C}_{\alpha\beta\lambda\mu}
= 2\Psi_4 V_{\alpha\beta}
V_{\lambda\mu} \,, \label{riwe}
\end{equation}
respectively.  Where the quantities $\Phi$, $\Psi_4$ and 
$V_{\alpha\beta}$ are given by
\[ \Phi\equiv H_{,\zeta\bar{\zeta}}\,,\hspace{6mm}
\Psi_4 = H_{,\bar{\zeta}\bar{\zeta}}\,,\hspace{6mm}
V_{\alpha\beta}\equiv 2\ell_{[\alpha}m_{\beta]} \,.\]
From (\ref{riwe}) we can see that the energy--momentum content
can correspond with vacuum, Einstein--Maxwell or pure 
radiation fields.  Moreover, the Petrov type is N, being 
$\vec{\ell}$ the repeated principal direction of the Weyl
tensor, which in fact is a constant vector field 
($\nabla_\alpha\ell_\beta=0$).

After some calculations, we have found that the Bel tensor 
$T^{\alpha\beta\lambda\mu}$ and our tensor $A^{\alpha\beta\lambda\mu}$ 
are 
\[ T^{\alpha\beta\lambda\mu}=4(\Phi^2+
\Psi_4\bar{\Psi}_4)\ell^\alpha\ell^\beta\ell^\lambda
\ell^\mu \,, \]
\[ \frac{1}{4}A^{\alpha\beta\lambda\mu}=
(3\Phi^2+\Psi_4\bar{\Psi}_4)\ell^\alpha\ell^\beta\ell^\lambda
\ell^\mu+
\ell^\alpha\ell^{(\beta}\nabla^\lambda\nabla^{\mu)}\Phi-
\ell^{(\beta}\ell^\lambda\nabla^{\mu)}\nabla^\alpha\Phi \]
\[\hspace{1cm} +\left[g^{\alpha(\beta}\ell^\lambda\ell^{\mu)}-
\ell^\alpha\ell^{(\beta}g^{\lambda\mu)}\right]\nabla^\sigma
\nabla_\sigma\Phi \,.\]
From this expression, the completely timelike component
is given by
\[ A(\vec{u}) = 4(3\Phi^2+\Psi_4\bar{\Psi}_4)
(\ell^\alpha u_\alpha)^4 \,. \]
Then, the vanishing of $A(\vec{u})$ implies the Minkowski 
space--time.

\vspace{4mm}

Finally, we are going to consider the Vaidya radiating 
space--time (see for instance \cite{KSHM}). For the sake
of brevity we only give here the completely timelike component
(\ref{comt}) of the tensor $A^{\alpha\beta\lambda\mu}$.
The line element of this spherically symmetric metric
can be written as follows
\[ ds^2=-2 F^2(u,v) du dv + r^2(u,v)(d\theta^2+\sin^2\theta 
d\varphi^2) \,,\]
where
\[ F^2(u,v)=f(u)\frac{\partial r}{\partial v} \,,
\hspace{8mm} \frac{\partial r}{\partial u}=\frac{1}{2}
f(u)\left(\frac{2m(u)}{r}-1\right) \,. \]
Here, $m(u)$ is the invariantly defined mass function.
As is well--known, the Petrov type of this metric is D.
Then, taking the following Newman--Penrose adapted basis
\[ \vec{\ell}=\frac{-1}{F}\frac{\partial}{\partial v}\,,
\hspace{5mm}
\vec{k}=\frac{-1}{F}\frac{\partial}{\partial u}\,,
\hspace{5mm}
\vec{m}=\frac{1}{\sqrt{2}r}\left(\frac{\partial}
{\partial \theta}+\frac{i}{\sin\theta}\frac{\partial}
{\partial \varphi}\right) \,, \]
where $\vec{\ell}$ and $\vec{k}$ are aligned with the principal
directions of the Weyl tensor, the Ricci tensor is given by
\[ R_{\alpha\beta}=2\Phi\ell_\alpha\ell_\beta\,,
\hspace{8mm} \Phi\equiv -\frac{m_{,u}}{r^2 r_{,u}} \,.\]
And the only non--zero component of the Weyl tensor in this
basis (see \cite{KSHM}) is
\[ \Psi_2 = -\frac{m(u)}{r^3(u,v)} \,,  \]
which in this case is real.

In terms of these quantities, we have found the following
expression for $A(\vec{u})$
\[ A(\vec{u}) = 2\left[\Psi_2+\Phi(\ell^\alpha u_\alpha)^2
\right]^2 + \]
\[ +\,4\Psi^2_2\left[36(\ell^\alpha u_\alpha)^2
(k^\beta u_\beta)^2-18(\ell^\alpha u_\alpha)(k^\beta u_\beta)
+1\right]+10\Phi^2(\ell^\alpha u_\alpha)^4 \,,\]
and again, taking into account that 
\[2(\ell^\alpha u_\alpha)(k^\beta u_\beta)\geq 1 \hspace{5mm} 
\Longrightarrow\hspace{5mm} 36(\ell^\alpha u_\alpha)^2
(k^\beta u_\beta)^2-18(\ell^\alpha u_\alpha)(k^\beta u_\beta)
+1\geq 1\,, \]
$A(\vec{u})$ vanishes if and only if the space--time is
the Minkowski space--time.
When we restrict ourselves to observers lying on the 2--planes
generated by the principal directions 
[$2(\ell^\alpha u_\alpha)(k^\beta u_\beta)=1$], which are 
precisely the observers that minimize $A(\vec{u})$, 
the result is 
\[ A(\vec{u}) = 2\left[\Psi_2+\Phi(\ell^\alpha u_\alpha)^2
\right]^2+ 4\Psi^2_2+10\Phi^2(\ell^\alpha u_\alpha)^4 \,.\]

\section*{Acknowledgments}
M.\'A.G.B. gratefully acknowledges the
{\it Comissionat per a Universitats i Recerca de la
Generalitat de Catalunya} for financial support.
C.F.S. gratefully acknowledges financial support in the
form of a fellowship from the Alexander von Humboldt
Foundation.

\appendix
\section{Useful definitions}
We first write down the Collinson tensor \cite{COLL}:
\begin{eqnarray}
T_{10}^{\alpha\beta\lambda\mu}=6 Q_{1}^{\alpha\beta\lambda\mu}+
Q_{2}^{\alpha\beta(\lambda\mu)}+Q_{3}^{\alpha\beta(\lambda\mu)} \, ,
\label{colli}
\end{eqnarray}
where
\begin{eqnarray*}
Q_{1}^{\alpha\beta\lambda\mu}&=&
R^{\lambda}_{\sigma}R^{\beta\mu\alpha\sigma}+
R^{\beta}_{\sigma}R^{\mu\alpha\lambda\sigma}+
\nabla^{\mu}\nabla_{\sigma}R^{\alpha\sigma\beta\lambda}-
\nabla^{\beta}\nabla_{\sigma}R^{\alpha\lambda\mu\sigma}-
g^{\alpha\beta}\nabla_{\rho}\nabla_{\sigma}R^{\mu\sigma\lambda\rho} \, ,\\
Q_{2}^{\alpha\beta\lambda\mu}&=&
-4\nabla^{\lambda}\nabla_{\sigma}R^{\alpha\sigma\beta\mu}-
6\nabla^{\mu}\nabla_{\sigma}R^{\alpha\beta\lambda\sigma}-
6g^{\alpha\mu}\nabla_{\sigma}\nabla^{\lambda}R^{\sigma\beta}+ \\
& &+6g^{\alpha\mu}\nabla_{\rho}\nabla_{\sigma}R^{\sigma\lambda\beta\rho}+
6\nabla^{\mu}\nabla^{\lambda}R^{\alpha\beta} \, , \\
Q_{3}^{\alpha\beta\lambda\mu}&=&
-8g^{\alpha\lambda}R_{\sigma\rho\hspace{2mm}\gamma}^{\hspace{3mm}\mu}
R^{\rho\gamma\beta\sigma}+
8R_{\sigma\hspace{3mm}\rho}^{\hspace{2mm}\lambda\mu}
R^{\beta\sigma\alpha\rho}+
8R^{\mu}_{\sigma}R^{\beta\sigma\lambda\alpha}+
8R_{\sigma\hspace{3mm}\rho}^{\hspace{2mm}\beta\mu}
R^{\lambda\sigma\alpha\rho}+ \\
& &+2R^{\mu}_{\sigma}R^{\beta\alpha\sigma\lambda}+
2R_{\sigma\hspace{3mm}\rho}^{\hspace{2mm}\mu\beta}
R^{\lambda\sigma\alpha\rho}+
3g^{\alpha\beta}R^{\mu}_{\sigma}R^{\lambda\sigma}-
g^{\alpha\beta}R_{\sigma\rho\hspace{2mm}\gamma}^{\hspace{3mm}\mu}
R^{\rho\gamma\lambda\sigma}\, ,
\end{eqnarray*}
which is divergence-free in the index $\alpha$ and
whose only symmetry on the indices $\beta$, $\lambda$ and $\mu$
is $T_{10}^{\alpha\beta[\lambda\mu]}-
T_{10}^{\alpha\mu[\beta\lambda]}=0$.

Next, in order to introduce other useful definitions,
recall the well--known decomposition of the
Riemann tensor into its irreducible parts under the
full Lorentz group:
\begin{eqnarray*} 
R_{\alpha\beta\lambda\mu}=C_{\alpha\beta\lambda\mu}+    
E_{\alpha\beta\lambda\mu}+G_{\alpha\beta\lambda\mu} 
\end{eqnarray*}
where $C_{\alpha\beta\lambda\mu}$ is the Weyl tensor and
\begin{eqnarray*}
E_{\alpha\beta\lambda\mu}&\equiv&
\frac{1}{2}\left(\tilde{R}_{\alpha\lambda}
g_{\beta\mu}-\tilde{R}_{\alpha\mu}g_{\beta\lambda}+
\tilde{R}_{\beta\mu}g_{\alpha\lambda}-
\tilde{R}_{\beta\lambda}g_{\alpha\mu}\right) \, , \;\;
\tilde{R}_{\alpha\beta}\equiv
R_{\alpha\beta}-\frac{1}{4}R g_{\alpha\beta} \, , \\
G_{\alpha\beta\lambda\mu}&\equiv&
\frac{R}{12}\left(g_{\alpha\lambda}
g_{\beta\mu}-g_{\alpha\mu}g_{\beta\lambda}\right) \, ,
\end{eqnarray*}
being $R\equiv R^{\mu}_{\mu}$ the scalar curvature.

The electric and magnetic parts of the Weyl tensor
associated to a timelike vector field $\vec{u}$ are:
\begin{eqnarray*}
E_{\alpha\lambda}(\vec{u})\equiv C_{\alpha\beta\lambda\mu}
u^{\beta}u^{\mu} \, , \; \;
H_{\alpha\lambda}(\vec{u})\equiv 
-\stackrel{*}{C}_{\alpha\beta\lambda\mu}
u^{\beta}u^{\mu} \, .
\label{weyl}
\end{eqnarray*}
These tensors are spatial (orthogonal to $\vec{u}$),
symmetric and traceless, and they fully determine
the Weyl tensor.

We can proceed analogously with the tensor
$E_{\alpha\beta\lambda\mu}$ and define (see \cite{BOSE}
for a more detailed study of these matters):
\begin{eqnarray*}
M_{\alpha\lambda}(\vec{u})\equiv
E_{\alpha\beta\lambda\mu}u^{\beta}u^{\mu}\, , \;\; 
N_{\alpha\lambda}(\vec{u})\equiv
-*E_{\alpha\beta\lambda\mu}u^{\beta}u^{\mu}\, .
\label{mater}
\end{eqnarray*}
We give here some of their properties, since they
are less known than the electric and magnetic
parts of the Weyl tensor:
\begin{eqnarray*}
M_{\alpha\lambda}=M_{\lambda\alpha}, 
\hspace{3mm} M_{\alpha\lambda}u^{\lambda}=0,
\hspace{3mm} M^{\mu}_{\mu}=
\tilde{R}_{\mu\nu}u^{\mu}u^{\nu}\, ,
\\
N_{\alpha\lambda}=-N_{\lambda\alpha},
\hspace{3mm}
N_{\alpha\lambda}u^{\lambda}=0,
\hspace{3mm} N^{\mu}_{\mu}=0 \, .
\end{eqnarray*}
The tensor $M_{\alpha\lambda}$ has 6 independent
components, while $N_{\alpha\lambda}$ has only 3.
Actually, they completely characterize the traceless
Ricci tensor:
\begin{eqnarray*}
\tilde{R}_{\alpha\beta}=
-2M_{\alpha\beta}-
4*N_{\sigma(\alpha}u^{\sigma}u_{\beta)}+
M_{\sigma}^{\sigma}
\left(g_{\alpha\beta}+2u_{\alpha}u_{\beta}\right) \, .
\end{eqnarray*}
From the previous definitions, it is clear that
\begin{eqnarray*}
N_{\alpha\beta}=R_{\alpha\beta}u^{\alpha}u^{\beta}=R=0 \;\;
\Longrightarrow \;\;
R_{\alpha\beta}=-2 M_{\alpha\beta} \, ,
\end{eqnarray*}
and this implies, in particular, that $R_{\alpha\beta}u^{\beta}=0$.

Let us finally give some formulae which is useful for
the derivations of some expressions in Sect. 4:
\begin{eqnarray*}
T^{\alpha\beta\lambda\mu}
u_{\alpha}u_{\beta}u_{\lambda}u_{\mu}&=&
E_{\alpha\beta}E^{\alpha\beta}+
H^{\alpha\beta}H_{\alpha\beta}+
M_{\alpha\beta}M^{\alpha\beta}+
N_{\alpha\beta}N^{\alpha\beta}+
\frac{R^{2}}{48} \, ,
\\
\left(\tilde{R}_{\alpha\rho}u^{\rho}\right)
\left(\tilde{R}^{\alpha}_{\sigma}u^{\sigma}\right)&=&
2N_{\sigma\rho}N^{\sigma\rho}-
\left(M_{\sigma}^{\sigma}\right)^2 \, ,\\
\tilde{R}_{\sigma\rho}\tilde{R}^{\sigma\rho}&=&
4M_{\sigma\rho}M^{\sigma\rho}-
4N_{\sigma\rho}N^{\sigma\rho} \, .
\end{eqnarray*}

\section{Electromagnetic case}
For space--times with an electromagnetic energy--momentum
content, Penrose and Rindler \cite{PERI} gave the following
modification of the Bel--Robinson tensor
(see \cite{PERI} for the spinor notations and conventions):
\begin{eqnarray}
t_{\alpha\beta\lambda\mu}&=&\Psi_{ABCD}
\overline{\Psi}_{A'B'C'D'}- 
\nonumber\\
& & -2\gamma \nabla_{CD'}\varphi_{AB}\nabla_{C'D}
\overline{\varphi}_{A'B'}+
6\gamma \nabla_{D(A'}\varphi_{(AB}\nabla_{C)|D'|}
\overline{\varphi}_{B'C')} \, ,
\label{penr}
\end{eqnarray}
where $\Psi_{ABCD}$ and $\varphi_{AB}$ are the Weyl
and the electromagnetic spinors respectively, and
$\gamma$ is the gravitational constant.
This tensor is symmetric and traceless in the first three
indices and has zero covariant derivative with respect
to the last one, provided that the Einstein field
equations hold:
\begin{eqnarray*}
\nabla^{\mu}t_{\alpha\beta\lambda\mu}=0 \, ,\\
t_{\alpha\beta\lambda\mu}=t_{(\alpha\beta\lambda)\mu}\, , \;\;
t^{\alpha}_{\alpha\lambda\mu}=0 \, .
\end{eqnarray*}
It is important to notice that the second part of this tensor is
formed with $(\nabla F)^2$ terms, so it cannot be expressed
by means of the Ricci tensor and its derivatives. Therefore
it is independent from the tensors considered above.

The tensorial expression for (\ref{penr}) is rather involved
but, in this case, it is useful in order to proof
a positivity property. It should be noted here that in
the final result we have returned to the initial signature metric 
$(-,+,+,+)$, and units such that $8\pi \gamma=c=1$.
With these conventions:
\begin{eqnarray}
t_{\alpha\beta\lambda\mu}&=&\frac{1}{4}
{\cal T}_{\alpha\beta\lambda\mu}
+2 \nabla_{(\alpha}F_{|\sigma|\beta}
\nabla_{\lambda)}F^{\sigma}_{\hspace{1mm}\mu}+
2 \nabla_{(\alpha}F_{|\sigma|\beta}\nabla^{\sigma}
F_{\lambda)\mu}- 
\nonumber \\
& & -\frac{1}{2}\nabla_{(\alpha}F_{|\sigma\rho|}
\nabla_{\beta}F^{\sigma\rho}g_{\lambda)\mu}+
\nabla_{\sigma}F_{\rho(\alpha}
\nabla^{\rho}F^{\sigma}_{\hspace{1mm}\beta}
g_{\lambda)\mu}-
g_{(\alpha\beta}\nabla^{\sigma}
F^{\rho}_{\hspace{1mm}\lambda)}
\nabla_{\rho}F_{\sigma\mu} \, ,
\label{penrose}
\end{eqnarray}
where ${\cal T}_{\alpha\beta\lambda\mu}$ is the
Bel--Robinson tensor and $F_{\alpha\beta}$ is the
electromagnetic tensor. From this
expression it is easily seen that (\ref{penrose})
satisfies the following positivity property:
\begin{eqnarray*}
t_{\alpha\beta\lambda\mu}
u^{\alpha}u^{\beta}u^{\lambda}u^{\mu}\;
&\geq& 0 \, , \;\; \forall \vec{u} \, , \;\;
u_{\mu}u^{\mu}<0 \, , \\
t_{\alpha\beta\lambda\mu}
u^{\alpha}u^{\beta}u^{\lambda}u^{\mu}
&=& 0 \Longleftrightarrow
\left\{ \begin{array}{l}
{\cal T}_{\alpha\beta\lambda\mu}
u^{\alpha}u^{\beta}u^{\lambda}u^{\mu}=0
\,\, \Longleftrightarrow \,\,
C_{\alpha\beta\lambda\mu}=0 \, ,\\
\dot{F}_{\lambda\mu}\equiv
u^{\alpha}\nabla_{\alpha}F_{\lambda\mu}=0  \, .
\end{array}
\right.
\end{eqnarray*}
To prove this, let us introduce the orthogonal projector
to $\vec{u}$, $h_{\alpha\beta}\equiv g_{\alpha\beta}
+u_{\alpha}u_{\beta}$. Then, we compute
$t_{\alpha\beta\lambda\mu}
u^{\alpha}u^{\beta}u^{\lambda}u^{\mu}$:
\begin{eqnarray*}
t_{\alpha\beta\lambda\mu}
u^{\alpha}u^{\beta}u^{\lambda}u^{\mu}&=&
\frac{1}{4}{\cal T}_{\alpha\beta\lambda\mu}
u^{\alpha}u^{\beta}u^{\lambda}u^{\mu}+
2 u^{\alpha}u^{\beta}
\dot{F}_{\sigma\alpha}
\dot{F}^{\sigma}_{\hspace{2mm}\beta}
+ \frac{1}{2}\left(
\dot{F}_{\sigma\rho}\dot{F}^{\sigma\rho}\right)= \\
&=& \frac{1}{4}{\cal T}_{\alpha\beta\lambda\mu}
u^{\alpha}u^{\beta}u^{\lambda}u^{\mu}+
\left(h^{\alpha\sigma}\dot{F}_{\alpha\beta}u^{\beta}\right)
\left(h_{\lambda\sigma}\dot{F}^{\lambda\mu}u_{\mu}\right)+\\
&+& \frac{1}{2}\left(h^{\alpha\sigma}h^{\beta\rho}
\dot{F}_{\alpha\beta}\right)
\left(h_{\lambda\sigma}h_{\mu\rho}
\dot{F}^{\lambda\mu}\right) \geq 0 \, ,
\end{eqnarray*}
and the equality holds only when $C_{\alpha\beta\lambda\mu}=0$ and
\begin{eqnarray*}
h^{\alpha\lambda}h^{\beta\mu}\dot{F}_{\lambda\mu}=0 \, ,
\; \;h^{\alpha\sigma}\dot{F}_{\alpha\beta}u^{\beta}=0
\Longleftrightarrow \dot{F}_{\alpha\beta}=0 \, .
\end{eqnarray*}

\end{document}